\documentstyle[aps,floats,epsfig]{revtex}

\begin{document} 
\preprint{} 
\draft 

\title{Ferroelectricity and structure of BaTiO$_3$ grown on 
YBa$_2$Cu$_3$O$_{7-\delta}$ thin films}

\author{Ch.~Schwan, F.~Martin, G.~Jakob, J.C.~Martinez, and H.~Adrian} 
\address{Institut f\"ur Physik, Johannes Gutenberg-Universit\"at Mainz, 
D-55099 Mainz, Germany} 

\date{\today} 

\maketitle 

\begin{abstract} 
We have investigated the crystal structure and the ferroelectric properties of 
BaTiO$_3$ thin films with YBa$_2$Cu$_3$O$_{7-\delta}$ as the bottom and Au as 
the top electrode. Epitaxial heterostructures of YBa$_2$Cu$_3$O$_{7-\delta}$ and 
BaTiO$_3$ were prepared by dc and rf sputtering, respectively. The crystal 
structure of the films was characterised by x-ray diffraction. The ferroelectric 
behaviour of the BaTiO$_3$ films was confirmed by hysteresis loop measurements 
using a Sawyer Tower circuit. We obtain a coercive field of 30~kV/cm and a 
remanent polarisation of 1.25~$\mu$C/cm$^2$. At sub-switching fields the 
capacitance of the films obeys a relation analogous to the Rayleigh law. This 
behaviour indicates an interaction of domain walls with randomly distributed 
pinning centres. At a field of 5~MV/m we calculate 3\% contribution of 
irreversible domain wall motion to the total dielectric constant. 
\end{abstract} 




\section{Introduction}
In recent years, there has been much interest in the preparation of 
superconducting field effect transistors (SuFETs) with a (Ba,Sr)TiO$_3$ thin 
film as gate insulator and YBa$_2$Cu$_3$O$_{7-\delta}$ (YBCO) as source-drain 
channel \cite{mannhart}. More recently the development of high frequency 
superconducting devices increased the interest of the community on 
YBCO/ferroelectric heterostructures \cite{jin}. From these studies it became 
clear that it is possible to grow high quality perovskite ferroelectric thin 
films on top of YBCO electrodes. This motivated us to investigate the room 
temperature ferroelectric and structural properties of YBCO/BaTiO$_3$/Au 
heterostructures.

The objective of this work is to analyse the domain wall contribution to the 
dielectric properties of high quality BaTiO$_3$ (BTO) thin films. So far, 
similar studies have only been done on Pt/PbZr$_x$Ti$_{1-x}$O$_3$ (PZT)/Pt 
heterostructures \cite{taylor}.

\section{Experiment}
The 100 nm thick YBCO layers and the BTO films were sputtered in-situ in pure 
O$_2$ atmosphere. The top Au electrodes were thermally evaporated through a 
shadow mask. Before deposition, the YBCO target was pre-sputtered for 1 hour and 
the BTO target for 30 minutes. After this time the dc-voltage of the plasma did 
not decrease anymore showing that the targets reached a stable state. For both 
processes the substrate target distance was about 20 mm. The heterostructures 
have been prepared on MgO substrates because of the better High Frequency 
properties of this material in comparison to SrTiO$_3$. The YBCO ground 
electrode was prepared following the method already described in details in Ref. 
\cite{schmitt}. The difference in the present work is that the post annealing 
was performed after the deposition of the BTO layer. The stoichiometric BTO 
targets were prepared at the University of Mainz. The BTO films were deposited 
by rf-sputtering in on axis geometry without a magnetron in O$_2$ pressures 
between 0.5 and 1.5 mbar. This was needed in order to prevent re-sputtering. 
During deposition, the rf power was regulated at 130 W resulting in a typical 
bias voltage of 30 V. The best films were deposited at 700 $^\circ$C. At the end 
of the process, the bi-layers were cooled down to room temperature during 1 hour 
and annealed in 1 bar oxygen pressure. The electrode contact size varied between 
0.5\ mm$^2$ and 1\ mm$^2$. X-ray diffraction patterns in Bragg-Brentano and 
four-circle geometry were obtained using a Philips X'Pert-MPD and a STOE STADI4 
diffractometer, respectively. The capacitance and the hysteresis loops were 
measured with a HP4284A LCR-meter and a self made Sawyer-Tower circuit, 
respectively. 

\section{Results and Discussion}
\subsection{X-ray diffraction}
\begin{figure}[p!] 
\centerline{\psfig{file=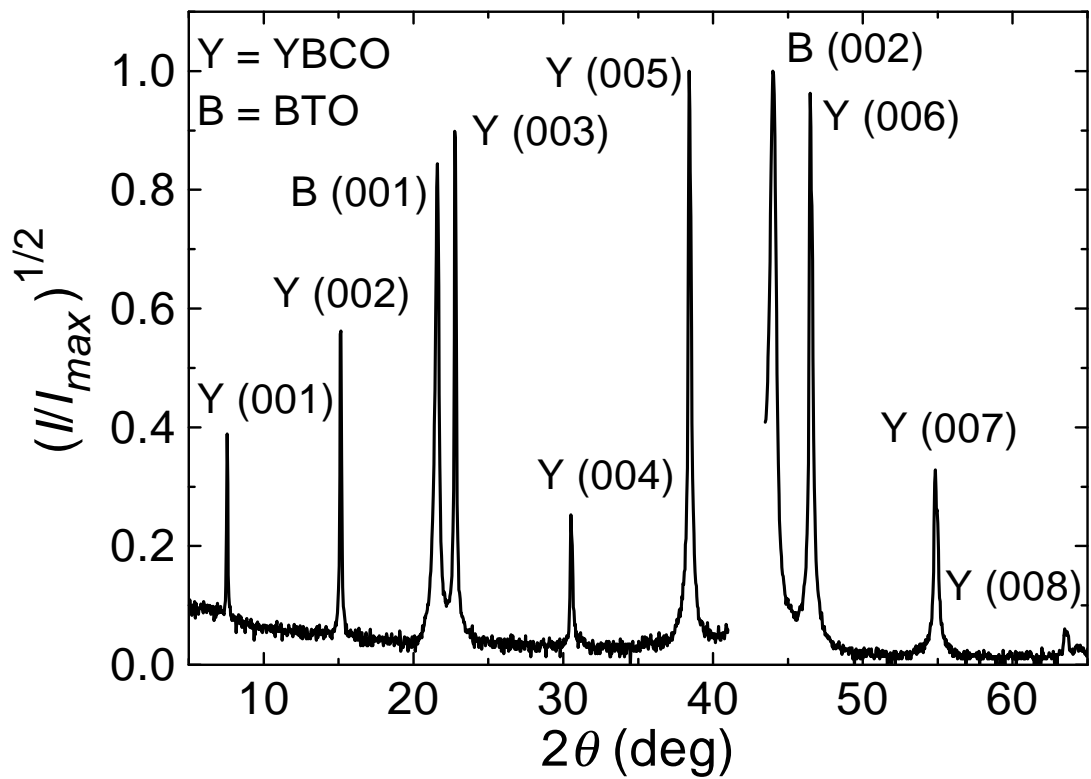,width=0.90\columnwidth}}
\vspace{0.5em} 
\caption{$\theta$/2$\theta$ scan of YBCO/BTO. Both YBCO and BTO show $c$-axis 
oriented growth. The YBCO peaks are
labelled with a Y and the BTO peaks with a B.} 
\label{theta} 
\end{figure}

The $\theta$/2$\theta$-scan shown in Fig. \ref{theta} reveals only $c$-axis 
oriented growth of the BTO films. However as shown by detailed TEM studies there 
may still exist small amounts of $a$-axis oriented domains embedded on a $c$-
axis matrix even if by x-ray diffraction no splitting caused by $a$-and $c$-axis 
oriented domains can be found \cite{Kim1}. The $c$-axis length of the BTO thin 
films in our work sputtered at 0.7 mbar is determined to be 3\% larger in 
comparison with the bulk material and coincides with the value of the work of 
Abe et al. \cite{Abe}. On the other hand, by using a deposition pressure of 1.5 
mbar we obtain the same $c$-axis length like the bulk material. Hence the 
lattice expansion at a lower pressure may result from atomic defects due to the 
high kinetic energy of the incident species. This explanation is in accordance 
with our observation that the lattice expansion occurs only for BTO but not for 
SrTiO$_3$ thin films. As barium ions are much heavier than strontium ions they 
can induce more serious damage upon impact on the previously deposited film. We 
expect therefore the lattice parameters to be also dependent on the deposition 
technique, as already reported by Srikant et al. \cite{Srikant}. 

\begin{figure}[p!] 
\centerline{\psfig{file=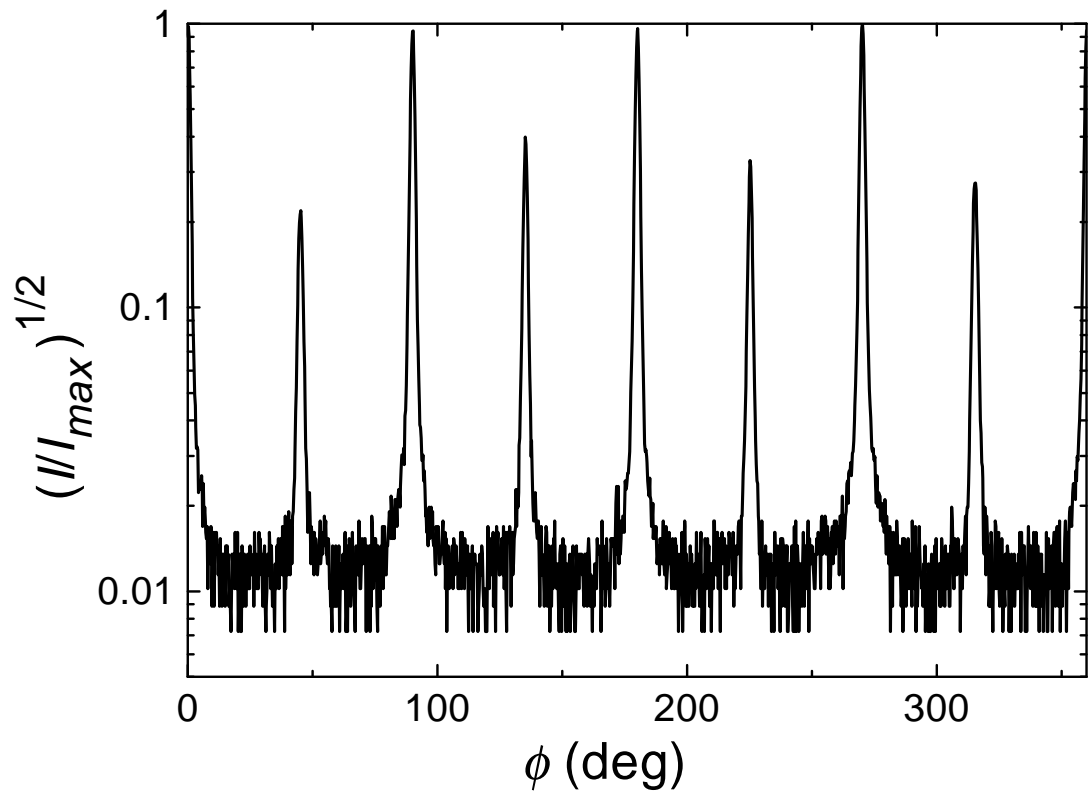,width=0.90\columnwidth}}
\vspace{0.5em} 
\caption{$\phi$ scan of the YBCO\,(103) reflection. There are 8 peaks indicating 
the alignment of YBCO parallel and
45$^\circ$ rotated to the MgO substrate. Both orientations exhibit a fourfold 
symmetry.} 
\label{phi} 
\end{figure}

\begin{figure}[p!] 
\centerline{\psfig{file=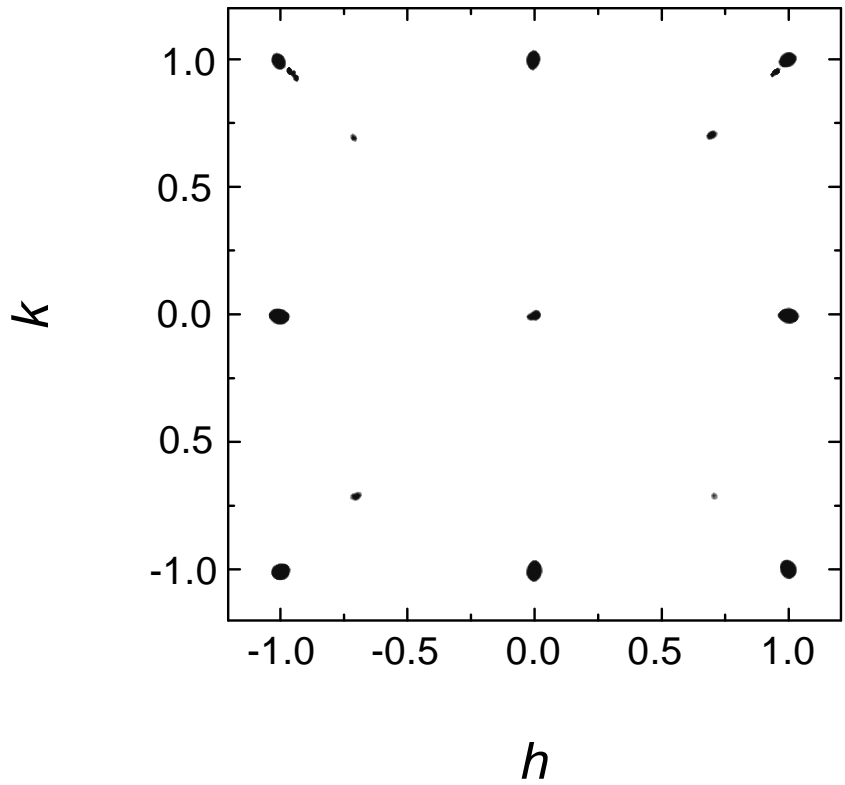,width=0.90\columnwidth}}
\vspace{0.5em} 
\caption{Two-dimensional scan of the plane $\ell=1$ for BTO. The 8 reflexes are 
caused by the epitaxial growth of BTO on
top of YBCO.} 
\label{qscan} 
\end{figure}

In order to characterise the in-plane orientation of the YBCO/BTO 
heterostructures we performed a $\phi$-scan of the YBCO (103) reflex (see Fig. 
\ref{phi}) and a two-dimensional scan of the plane $\ell=1$ for BTO (see Fig. 
\ref{qscan}). For both films there are 8 peaks, indicating a fourfold symmetry 
and an in-plane alignment of YBCO's and BTO's crystallographic axes parallel and 
rotated 45$^\circ$ relative to the MgO lattice. The rotated alignment can be 
explained by the fact that 2 lattice constants of MgO equals 3 diagonal lattice 
parameters of YBCO. Since BTO grows on YBCO epitaxial it reproduces this 
45$^\circ$ rotation. 

A fourfold symmetry was obtained by the use of SrTiO$_3$ substrates. The 
orientation matrix gives for our films the in-plane lattice constants 
$a=b=(0.3994\pm 0,0002)$\ nm for BTO. This value is in excellent agreement with 
the result on bulk materials. For YBCO we measured $a=0.3847$ nm, $b=0.3851$ nm. 
These values coincide well with the average between $a$ and $b$ in bulk 
materials. This is due to the usual mixing of these two orientations observed in 
thin films done on standard substrates. The results above show, that the $c$-
axis distortion of BTO cannot be explained by the requirement to conserve the 
unit cell volume, according to the Poisson ratio. This indicates that the high 
energy of barium at the sputtering process might be indeed responsible for the 
observed lattice distortion. 

The mosaic spreads of the different layers were determined by measuring the 
rocking curves of BTO (002) and YBCO (005). For BTO we obtain a FWHM (full width 
at half maximum) of 0.38$^\circ$ and for YBCO 0.32$^\circ$, indicating a good 
epitaxial growth of the heterostructures. The small FWHM obtained in our BTO 
films can be explained by a reduced lattice mismatch between YBCO and BTO. For 
comparison, Wills et al. got for $a$-axis oriented BTO on LaAlO$_3$ substrates a 
FWHM of 0.6$^\circ$ \cite{Wills}. 

\subsection{Ferroelectric hysteresis loops}
\begin{figure}[p!]
\centerline{\psfig{file=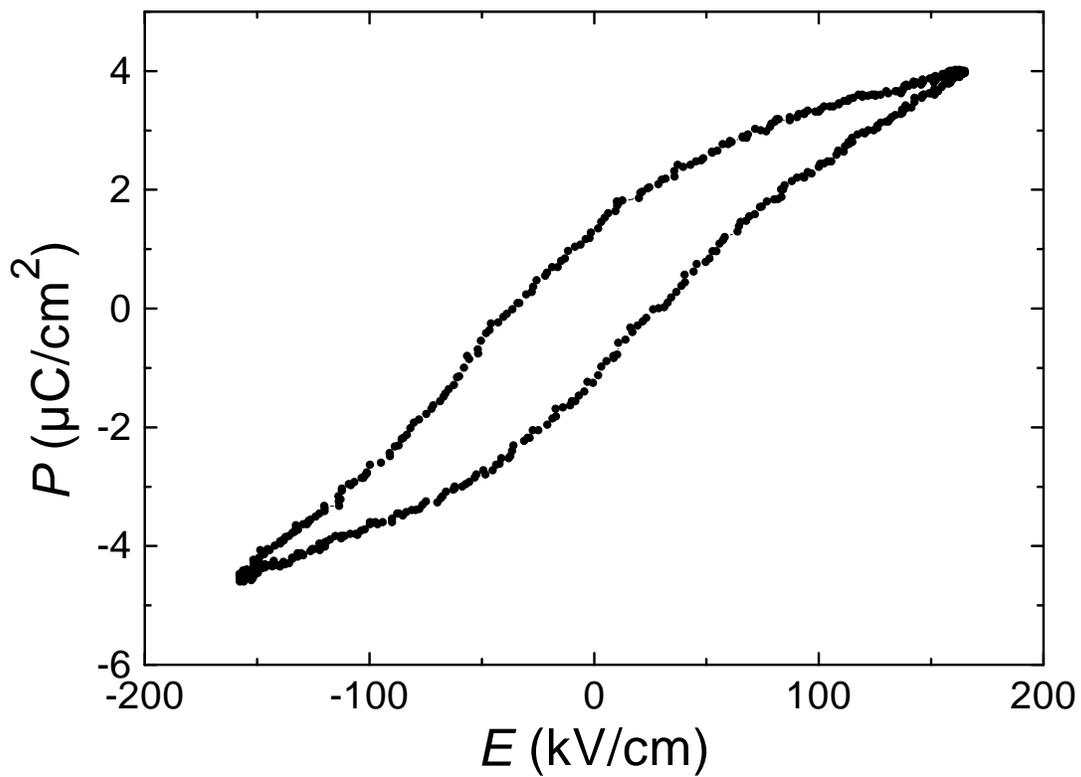,width=0.90\columnwidth}}
\vspace{0.5em} 
\caption{Polarisation vs field hysteresis loop. Using a Sawyer-Tower circuit, we 
obtain a remanent polarisation of
$P_{\rm r}=1.25 \mu$C/cm$^2$ and a coercive field of $E_{\rm c}=30$ kV/cm.} 
\label{Hysteresis}
\end{figure}

The non-remanent component of polarisation is caused by non-ferroelectric ionic 
and electronic polarisability as well as field induced reversible ferroelectric 
domain wall motion. The remanent component is induced by switching ferroelectric 
domains. Using a Sawyer-Tower circuit, we obtain a remanent polarisation of 
$P_{\rm r}=1.25 \mu$C/cm$^2$ and a coercive field of $E_{\rm c}=30$ kV/cm (Fig. 
\ref{Hysteresis}). The remanent polarisation is consistent with other results 
obained by rf-sputtering \cite{Song,Cheng}. But Yoneda et al. even observed for 
30 monolayers thick BTO films a ferroelectric hysteresis up to 600 $^\circ$C 
\cite{Yoneda}. Measuring the dielectric constant of our 300 nm thick BTO films, 
from room temperature to 200 $^\circ$C, did not indicate any sign of a 
ferroelectric phase transition, as well. This result is rather surprising since 
bulk BTO is known to have a Curie temperature of about 120 $^\circ$C 
\cite{jona}. This increase in the curie temperature could be due to internal 
stresses induced by the substrate.

\subsection{Contribution of the domain walls to the permittivity} 
\begin{figure}[p!] 
\centerline{\psfig{file=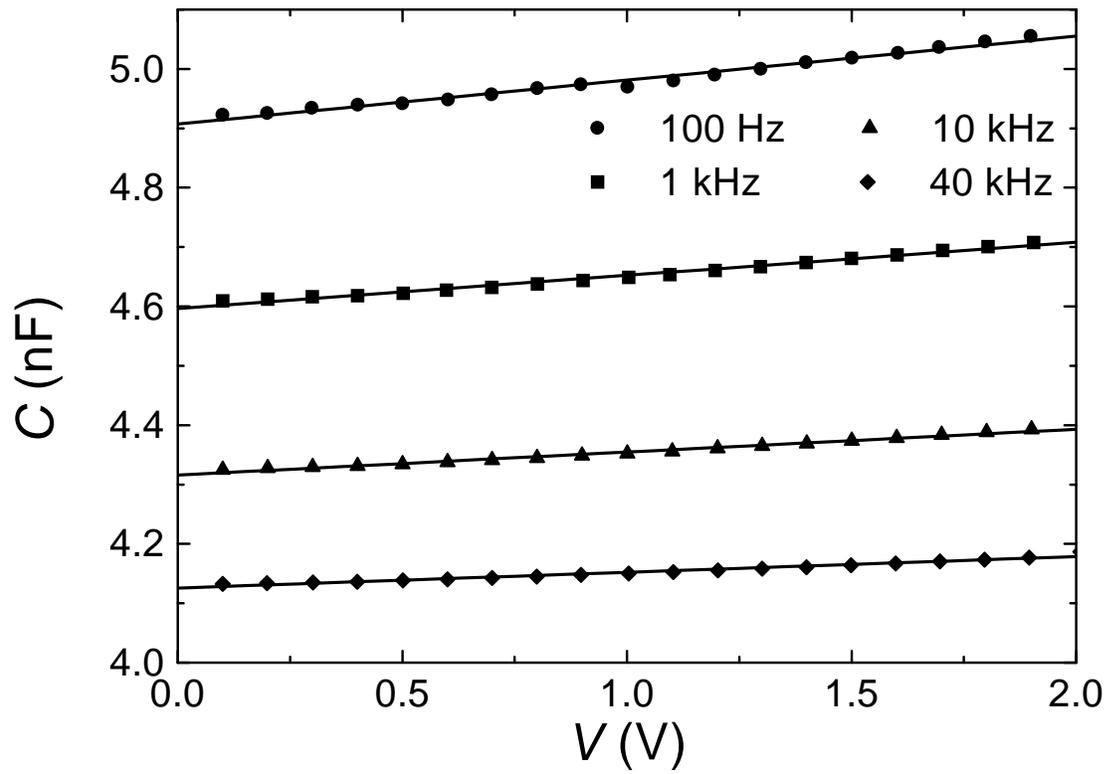,width=0.90\columnwidth}}
\vspace{0.5em} 
\caption{Field dependence of the capacitance at different frequencies. The full 
lines represent the fits according to the
Rayleigh law.} 
\label{ACField} 
\end{figure}

\begin{figure}[p!] 
\centerline{\psfig{file=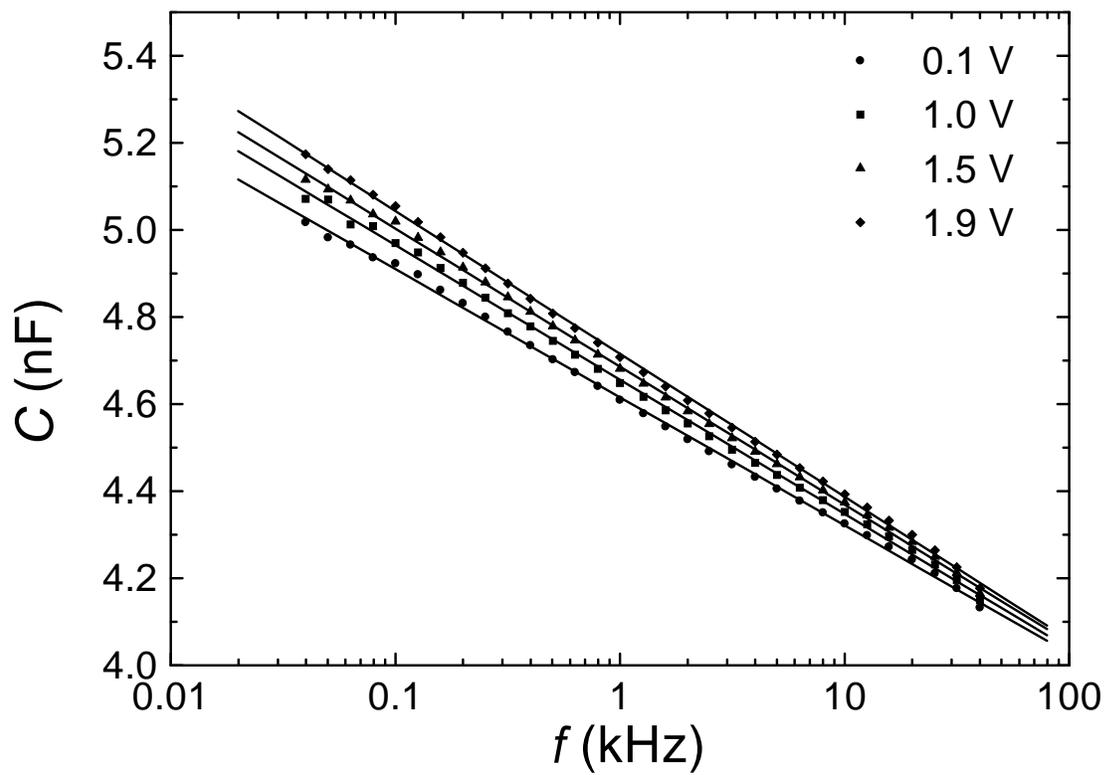,width=0.90\columnwidth}}
\vspace{0.5em} 
\caption{Frequency dependence of the capacitance. The full lines represent 
logarithmic fits.} 
\label{Frequency}
\end{figure}

\begin{figure}[p!] 
\centerline{\psfig{file=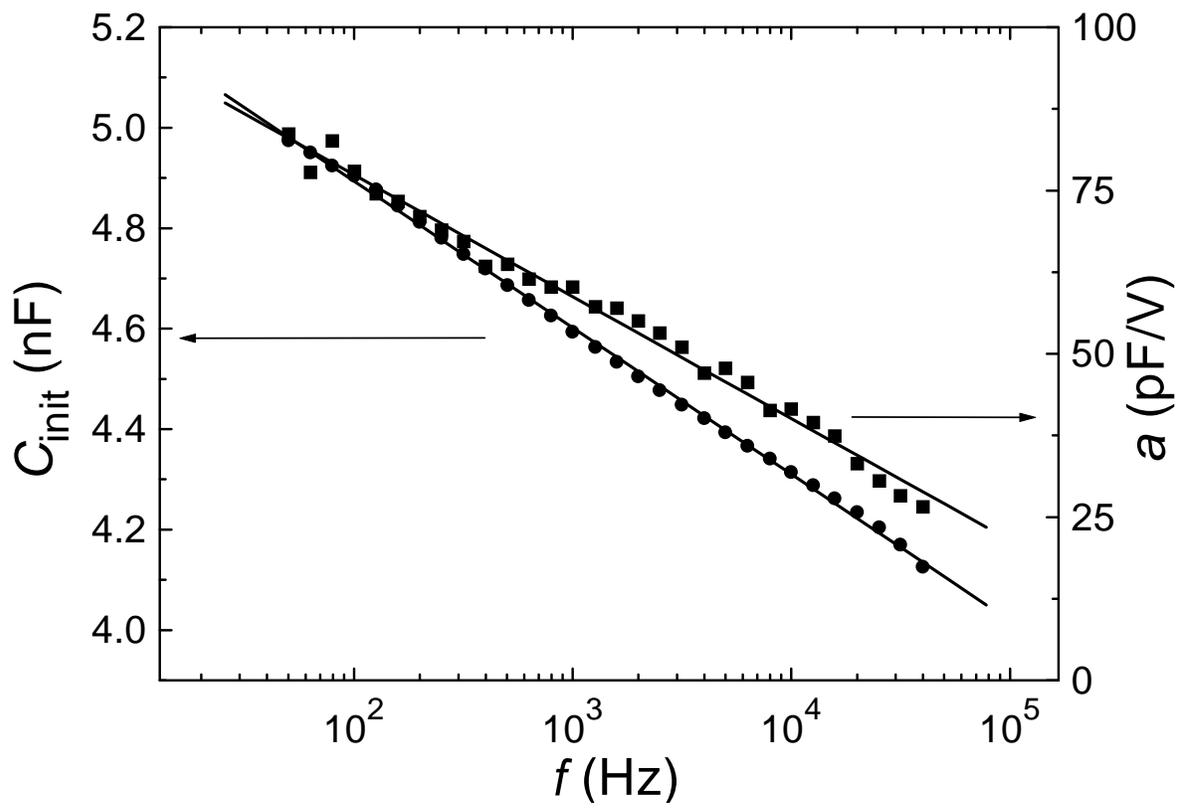,width=0.90\columnwidth}}
\vspace{0.5em} 
\caption{Frequency dependence of the reversible and irreversible parameters 
$C_{\rm init}$ and {$a$} of the capacitance.}
\label{Parameter} 
\end{figure}

\begin{figure}[p!] 
\centerline{\psfig{file=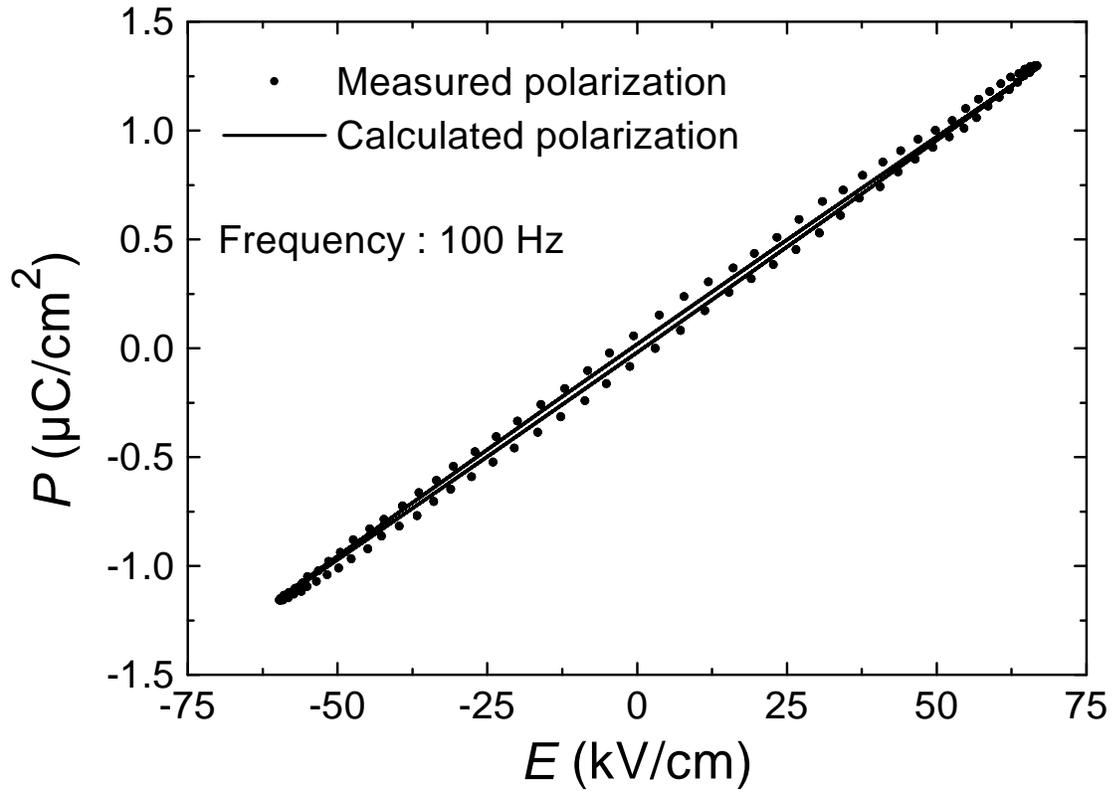,width=0.90\columnwidth}}
\vspace{0.5em} 
\caption{$P(E)$ hysteresis loop for $E_{\rm 0}=2$ MV/cm at 100 Hz. Circles 
correspond to experimental data and the full
lines are calculated with $C_{\rm init}$ and {$a$} extracted from Fig. 
\ref{ACField}.} 
\label{RayleighHys} 
\end{figure}

The major research in ferroelectric thin films emphasises the nucleation, growth 
and switching of domains because these features are very important for 
nonvolatile memory applications. In contrast, contributions to the dielectric or 
the piezoelectric constant at sub-switching fields due to domain wall 
displacements and vibrations were investigated mainly for single crystals or 
ceramics. We follow here the procedure of Taylor et al. by extending the models 
for the interaction and pinning of domain walls with randomly distributed 
pinning centres in magnetic materials to the dielectric displacement in 
ferroelectric thin films \cite{taylor}. Below we will address the dielectric 
response of a BTO thin film both in terms of the weak-field (Rayleigh law) and 
the logarithmic frequency dependence.\ The Rayleigh law is valid for low-field 
conditions, where no nucleation of domains occur and the average structure of 
the domain walls drifts within the sample as the field is cycled 
\cite{Damjanovic2}. In order to achieve the linear Rayleigh behaviour for the 
dielectric constant we cycled the sample about 10$^7$ times prior to the 
measurement with an amplitude below the coercive voltage. Figure \ref{ACField} 
shows the linear dependence of the capacitance versus ac voltage amplitude 
$V_0$. 
From the capacitance at 100 Hz, and from the geometry of our electrodes we 
deduce a dielectric constant $\epsilon=300$. 
In accordance to the Rayleigh law the weak-field dependence can be described by 
the formula $C=C_{\rm init}+aV_0$. $C_{\rm init}$ represents the intrinsic 
lattice and reversible domain wall movement. The second term $aV_0$ with the 
Rayleigh coefficient $a$ is responsible for the weak-field hysteresis due to the 
pinning and depinning of domain walls. According to N\'{e}el it is caused by the 
irreversible displacement of the walls between two meta-stable states when the 
applied field is large enough to overcome the barrier.\ In Fig. \ref{Frequency} 
we show the logarithmic dependence of the capacitance on frequency. As for 
magnetic systems this logarithmic behaviour is generally attributed to domain 
wall motion across randomly distributed pinning centres \cite{taylor, Natterman}. 
In order to determine whether the reversible $C_{\rm init}$ or the irreversible 
Rayleigh coefficient $a$ is responsible for this behaviour, we extracted them 
for all frequencies from the ac field dependence of the capacitance. The result 
indicating the logarithmic frequency dependence of both parameters is presented 
in Fig \ref{Parameter}. Using a Sawyer-Tower circuit we finally measured the 
non-saturating hysteresis loop (Rayleigh loop) of our samples at sub-switching 
fields (see Fig. \ref{RayleighHys}). The full line representing the calculated 
hysteresis curve is based on the LCR-meter measurement by using the formula 
$P=(\epsilon_{init}+\alpha E_0)E\pm \alpha /2(E_0^2-E^2)$, where $+$ stands for 
the decreasing and $-$ for the increasing field. $\epsilon_{\rm init}$ and 
$\alpha$ are derived from $C_{\rm init}$ and $a$ respectively. $E_0$ designates 
the maximum field reached during the cycle. While the slopes of the loops 
coincide, the area of the calculated hysteresis curve is underestimated in 
comparison to the measured one. This fact suggests that the observed hysteresis 
is not only due to the Rayleigh-type irreversible contribution of the dielectric 
constant, but by the losses having their origin in the series resistance and 
parallel conductivity of the capacitor. This explanation is confirmed by the 
fact that the calculated curve is only based on the imaginary and not on the 
real part of the impedance.

So far there is no universal behaviour for the frequency response of 
capacitors based on insulating perovskites. While Taylor et al. observed, for 
Pt/PZT/Pt capacitors, in the range from 10$^{-1}$ Hz to 10$^4$ Hz an exponential 
law between capacitance and frequency \cite{taylor}, Zafar et al. used for 
Pt/(Ba,Sr)TiO$_3$/Pt between 10$^2$ Hz and 10$^6$ Hz a power law to describe 
their data \cite{Zafar}. 

While the logarithmic frequency dependence is in accordance to the contribution 
of domain walls, the power law dependence is known as the Curie-von Schweidler 
law. Obviously further studies are necessary to determine the correct 
relationship between the dielectric constant and the frequency. Maybe our data 
and the results of Taylor et al, fit better to the logarithmic frequency 
dependence because the films exhibit a remanent polarisation and thus a higher 
domain wall contribution in comparison to non-ferroelectric (Ba,Sr)TiO$_3$ 
films. The relative irreversible domain wall contribution to the 
total capacitance can be obtained by the ratio $aV/C$. If we apply an sinusoidal 
electrical field with a frequency of 100 Hz and an amplitude of $E_0=5$ MV/cm we 
obtain a ratio of 3\%. This value is much smaller than the 21\% calculated by 
Taylor et al. for 1.3 $\mu$m thick PZT thin films. Since the irreversible part 
of the dielectric constant determines the remanent polarisation it is reasonable 
for BTO to obtain a lower contribution of irreversible domain wall movement than 
for PZT because the remanent polarisation of PZT thin films is typically in the 
range of 40 $\mu$C/cm$^2$. This is more than one magnitude larger in comparison 
with our BTO thin films, revealing a remanent polarisation of 1.25 
$\mu$C/cm$^2$.\ Damjanovic et al. determined the Rayleigh parameters describing 
the piezoelectric response of PZT and BTO ceramics by applying an external ac 
pressure to the samples \cite{Damjanovic1,Damjanovic2,Damjanovic3}. The Rayleigh 
law for the piezoelectric behaviour can be described by the formula $d=d_{\rm 
init}+\beta X_0$ with the piezoelectric constant $d$, the amplitude $X_0$ of the 
applied pressure and the Rayleigh coefficient $\beta$. In contrast to our 
results for thin films they observed, for the coarse grain (average grain size 
27 $\mu$m) BTO ceramic, a higher relative irreversible contribution $\beta X/d$ 
than for soft PZT. Thus the potential barrier for a domain wall displacement is 
not dominated by the material, but by the grain size and the type of domain walls 
because 180$^\circ$ domain walls influence the dielectric constant but not the 
piezoelectric coefficient \cite{Zhang}. Demartin concluded for fine grain 
samples (0.7 $\mu$m grain size) of BTO that domain walls are clamped to a 
considerable degree by the high internal pressure \cite{Demartin} and hence they 
contribute less to the piezoelectric coefficient. In contrast to coarse grain 
ceramics (26 $\mu$m grain size), the domain walls created in fine grain ceramics 
are not sufficient to relieve the stress produced by the phase transition, 
resulting in a high internal stress and a weak activity of the domain walls. The 
decrease in dielectric constant with decreasing grain size (from 50 nm to 12 nm) 
arises also from the absence of 180$^\circ$ domain boundaries in smaller grains 
\cite{Jang,Kim2}. On the other hand, it is well known that polycrystalline BTO 
with a grain size of approximately 1 $\mu$m possesses a much higher permittivity 
at room temperature than a single crystal BTO due to the higher contribution of 
domain walls \cite{Arlt}. The much larger film thickness of 1.3 $\mu$m chosen by 
Taylor and Damjanovic \cite{taylor} yields an enhanced activity of the domain 
walls and a higher Rayleigh coefficient in comparison to our 300 nm thick BTO 
films. 

The grain size and the stresses due to the phase transition determine the  
dielectric constant in ceramics. But ferroelectric thin films are additionally 
influenced by the existence of interfaces and additional stresses, caused by the 
lattice mismatch between film and substrate.

\section{Conclusion}

The small lattice mismatch and chemical compatibility between YBCO and BTO allow 
a good crystal structure for heterostructures of both materials. X-ray 
diffraction patterns display a high crystalline order of both YBCO and BTO 
whereas their in-plane orientation is influenced by the choice of MgO as 
substrate material. The sub-switching ac electric field dependence of the 
permittivity may be described in terms of the Rayleigh law. At a field of 5 MV/m 
the total dielectric constant contains 3\% contribution of irreversible domain 
wall movement. In addition to $C_{init}$ the irreversible Rayleigh parameter 
representing irreversible domain wall movement shows also a linear logarithmic 
frequency dependence of permittivity. This behaviour can be explained by pinning 
of domain walls. In comparison to conventional Pt as base electrode for 
ferroelectric memories, YBCO offers the advantages of better fatigue behaviour.


The authors gratefully acknowledge support by the Materialwissenschaftlichen 
Forschungszentrum (MWFZ), Mainz, the Training and Mobility of young researchers 
(TMR) program of the European Union under grant number ERBFM-BICT972217 and the 
German Bundesministerium f\"{u}r Bildung, Wissenschaft, Forschung und Technologie 
(BMBF) under project number 13N6916.

\newpage

\newpage

\newpage
\newpage

\newpage

\newpage

\newpage

\newpage

\end{document}